\documentclass[manuscript]{aastex} 
\usepackage{txfonts} 
\usepackage[english]{babel} 
\usepackage{t1enc} 
\usepackage[ansinew]{inputenc} 
\usepackage{graphicx} 
\usepackage{subfig} 
\usepackage{url}

\begin{document} 
 
\title{The VLBA Calibrator Search for the BeSSeL Survey} 
\shorttitle{VLBA Calibrator Search} 

\author{K. Immer \altaffilmark{1,2}, A. Brunthaler \altaffilmark{1}, M. J. Reid \altaffilmark{2}, \\
A. Bartkiewicz \altaffilmark{3}, Y. K. Choi \altaffilmark{1}, K. M. Menten  \altaffilmark{1}, L.
Moscadelli \altaffilmark{4}, A. Sanna \altaffilmark{1}, Y. W. Wu \altaffilmark{5}, Y. Xu \altaffilmark{5}, 
B. Zhang \altaffilmark{1}, X. W. Zheng \altaffilmark{6}}
\email{kimmer@mpifr-bonn.mpg.de}

\altaffiltext{1}{Max-Planck-Institut f$\ddot{\textnormal{u}}$r Radioastronomie, Auf dem H$\ddot{\textnormal{u}}$gel 69, 53121 Bonn, Germany}
\altaffiltext{2}{Harvard-Smithsonian Center for Astrophysics, 60 Garden Street, 02138 Cambridge, MA, USA}
\altaffiltext{3}{Toru$\acute{\textnormal{n}}$ Centre for Astronomy, Nicolaus Copernicus University, Gagarina 11, 87-100 Toru$\acute{\textnormal{n}}$, Poland}
\altaffiltext{4}{Osservatorio Astrofisico di Arcetri, INAF, Largo E. Fermi 5, 50125 Firenze, Italia}
\altaffiltext{5}{Purple Mountain Observatory, Chinese Academy of Sciences, 2 West Beijing Road, Nanjing 210008, P.R. China}
\altaffiltext{6}{Department of Astronomy, Nanjing University, 22 Hankou Road, Nanjing 210093, P.R.China}

\keywords{Galaxy: general --- HII regions --- astrometry --- surveys --- radio continuum}

\begin{abstract}
We present the results of a survey of radio continuum sources near the 
Galactic plane using the Very Long Baseline Array (VLBA).  Our observations are designed to identify compact
extragalactic sources of milliarcsecond size that can be used for parallax measurements in the 
Bar and Spiral Structure Legacy Survey.  We selected point sources from
the NVSS and CORNISH catalogs with flux densities above 30 mJy and within $1.5\degr$ 
of known maser targets.  Of the 1529 sources observed, 199 were detected.
For sources detected on 3 or more baselines, we determined accurate positions and 
evaluated their quality as potential calibrators.  Most of the 1330 sources 
that were not detected with the VLBA are probably of extragalactic origin. 
\end{abstract}

\section{Introduction} 
The Very Long Baseline Array (VLBA) calibrator search is a preparatory survey for the Bar and 
Spiral Structure Legacy (BeSSeL) survey \citep{Brunthaler2011} which will study the spiral 
structure and kinematics of the Milky Way. The BeSSeL survey will determine 
distances, via trigonometric parallax, and proper motions for methanol and 
water masers at 6.7 and 22 GHz in several hundred high mass star forming regions in our Galaxy. 
These results will allow us to locate spiral arms and determine 
the 3-dimensional motions of massive, young stellar objects in the Galaxy.
(for further information about the BeSSeL survey see 
\url{http://www.mpifr-bonn.mpg.de/staff/abrunthaler/BeSSeL/index.shtml})

Currently the VLBA calibrator search list contains information on compact
radio sources from the surveys of \citet{Ma1998,Beasley2002,Fomalont2003,
Fey2004,Petrov2005,Petrov2006,Petrov2008,Kovalev2007}.
Recently \citet{Petrov2011} published the results of a survey 
of 489 calibrator sources close to water masers in the Galactic plane 
for the Japanese VLBI Exploration of Radio 
Astrometry\footnote{\url{http://veraserver.mtk.nao.ac.jp/}} (VERA) project.
However, a search for background sources near any given maser target
with the VLBA calibrator search tool usually yields a source
within about $3\degr$ on the sky.  Since transfering interferometer
phases over such a large angular distance often provides poor results, such
background sources are not always preferable for accurate parallax measurements.
Since systematic errors scale approximately linearly with the separation
between target and calibrator source, it is valuable to find background
sources as close to the target as possible.    

This paper presents the results from our calibrator survey.
While this survey was tailored for finding compact extragalactic sources
for the BeSSeL parallax observations, it should also be useful
for other VLBA observations.  Also, the compact sources found in
this survey should be useful for future EVLA and ALMA observations.
Since, historically, there has been a shortage of good calibrators near 
the Galactic plane, this survey should be especially valuable for 
interferometric observations of Galactic sources.

\section{Source Selection and Observation}

Point-like sources from the NVSS (sizes $<20\arcsec$) and CORNISH catalogs 
\citep[provided by][]{Condon1998,Purcell2008} with flux densities above 30 mJy 
and within circles with a radius of $1.5\degr$ around the 109 target maser sources
formed the survey list. Since a Bessel parallax observation benefits from at least one
calibrator with a milliarcsecond-accurate position, we added
between 1 and 3 sources from the ICRF2\footnote{International Celestial 
Reference Frame} catalog \citep{Fey2009} that were within $5\degr$ of
each target maser position.
Since some target masers were separated by less than
$3\degr$, a number of candidates were observed 2 or more times and the total
areal coverage was $\sim600$ square degrees.  However, the calibrator
survey is {\bf not} complete to 30 mJy, since time restrictions
typically allowed observing only the $\approx25$ most promising
sources for each target position. Thus, in some search fields, 
a small number of the weaker and more distant sources were not observed.          

\begin{figure} 
\centering 
\includegraphics[width=16.5cm]{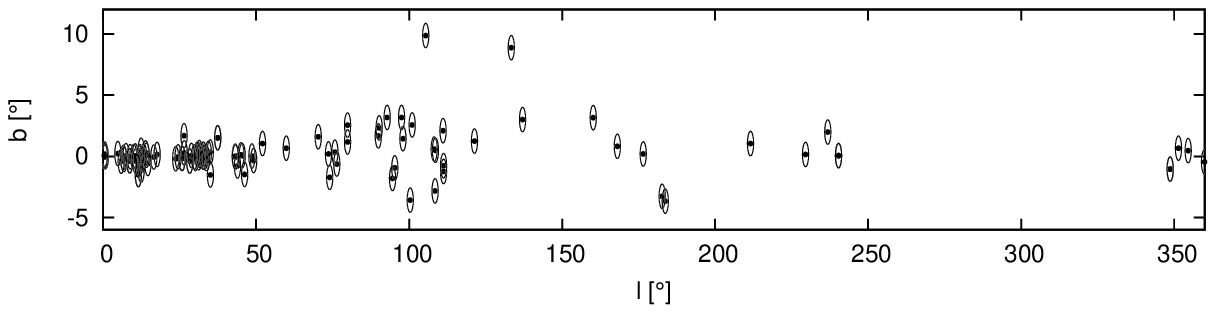} 
\caption{Areal coverage of the survey with Galactic longitude l and Galactic latitude b. 
The black dots mark the positions of the target maser sources, the black ellipses show 
the circular area on the sky over which radio point sources were selected as 
candidate calibrator sources.\label{GalLatLong}}
\end{figure}

In total, 1529 sources were selected for observations with the VLBA.
Figure \ref{GalLatLong} shows the covered area of the survey in 
Galactic coordinates. The black dots mark
the positions of the target masers, the black ellipses represent the
circular area on the sky searched for calibrators.                                
All sources were observed at X-band in left-circular polarization.
We used eight 16-MHz bands, with band center frequencies spanning the range 
from 8109.49 to 8583.49 MHz.
The bands were spaced in a minimum redundancy configuration to sample,
as uniformly as possible, all frequency differences in order to minimize
multi-band delay sidelobes. Each survey source was observed for three minutes. 
Three strong fringe-finders were observed at the beginning and the
end of the typically five-hour tracks.

The data reduction was conducted with the National Radio Astronomy Observatory's\footnote{The National 
Radio Astronomy Observatory (NRAO) is a facility of the 
National Science Foundation operated under cooperative agreement by 
Associated Universities, Inc.}
Astronomical Image Processing System, using the scripting software ParselTongue\footnote{http://www.radionet-eu.org/rnwiki/ParselTongue}
\citep{Kettenis2006}.
In the first step, we corrected the data for the latest Earth Orientation Parameter (EOP) values. 
With the AIPS task TECOR, we derived corrections for the ionospheric Faraday rotation from 
maps of total electron content, which are available in the CDDIS\footnote{NASA's Crustal Dynamics Data
Interchange System} data archive. 
Then, one scan on a strong fringe-finder was used to solve for single-band
delays and to align the phases of all IFs. To remove clock offsets, mini-geodetic block sources were fringe fitted.
Next, we determined multi-band delays and fringe-rates for the fringe-finder blocks placed 
at the beginning and end of the observations. We averaged these values for each block and 
removed their effects from the survey data. Afterwards, the residual multi-band delays and 
fringe rates for the candidate sources were modeled as owing to source position offsets and 
revised positions were determined by least-squares fitting. In addition, we generated plots of 
interferometer visibility amplitude versus the baseline length by vector averaging 
the data over each scan for each interferometer baseline in order to 
determine the source compactness.

\section{Detections}
 
199 candidate sources were detected on at least one VLBA baseline, of which 70 
are sources of the ICRF2 catalog and 30 were previously detected by \citet{Petrov2011}.
Our goal was to determine reliable positions for the candidate calibrator sources.
For 25 of the ICRF2 sources, the ICRF2 catalog position was significantly better 
than our measured source position. For these sources, we replaced the source coordinates by an average
of the measured and the ICRF2 coordinates. 

Depending on a source's compactness, a grade from
A to D was assigned, representing their quality as
a potential calibrator.  Compactness was determined by examining the
visibility plots by eye and estimating the maximum baseline length
(in wavelength units) for which the normalized fringe visibility was
greater than 0.2.  For grade A sources, which
are the most compact, the normalized fringe visibility falls
below this threshold at $>200$ M$\lambda$.  Grade B and C
sources are defined to have normalized visibilities greater than
0.2 for baselines longer than 100 and 50 M$\lambda$, respectively. 
Grade D sources were more resolved than grade C sources, but still
detected on 3 or more baselines.                                 
Finally, grade F sources were detected on only one or two baselines. 
The two shortest VLBA baselines have lengths of 236 km (PT--LA) and
417 km (KP--PT).  Grade F sources were not used in the following 
statistical analysis.                                                              

\begin{figure} 
	  \centering 
	  \subfloat[J0256+6827 - Grade A source]{\includegraphics[width=8cm]{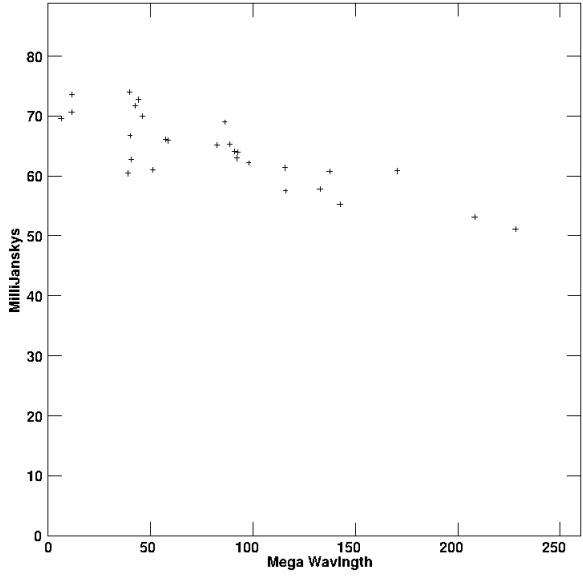}}	
	  \subfloat[J1825-1718 - Grade B source]{\includegraphics[width=8cm]{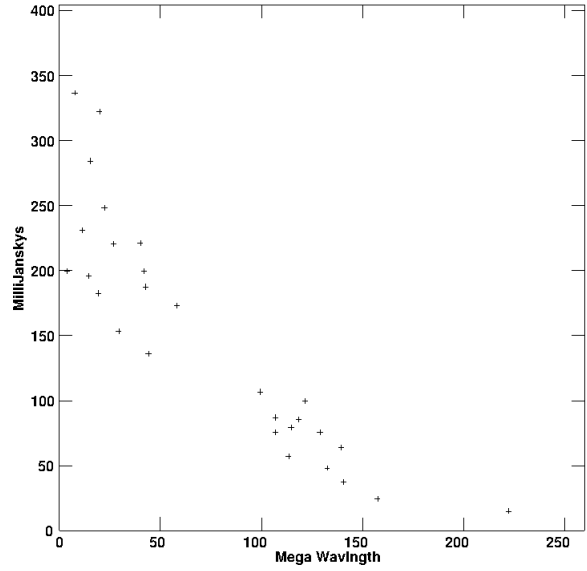}}\\
	  \subfloat[J1751-2523 - Grade C source]{\includegraphics[width=8cm]{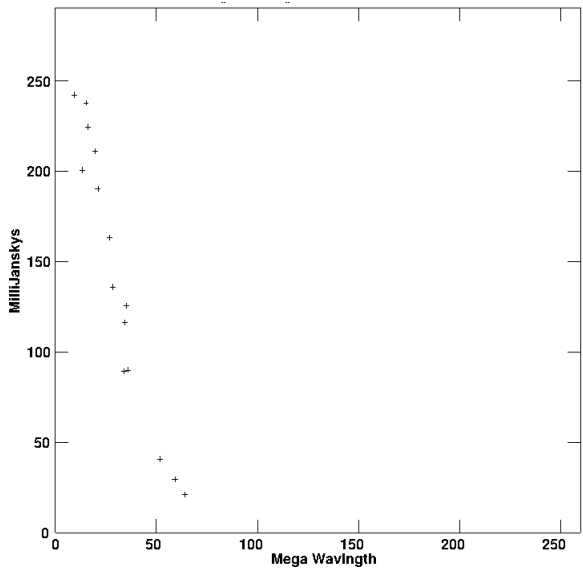}}
	  \subfloat[J1758-2343 - Grade D source]{\includegraphics[width=8cm]{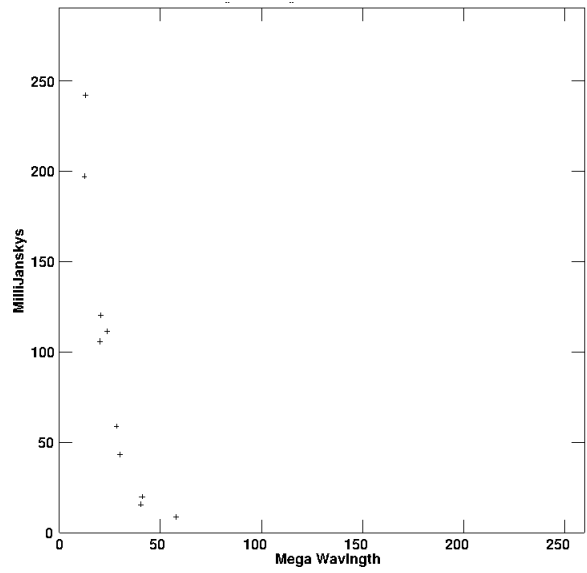}}	
		\caption{Examples of the visibility amplitude versus baseline length
plots for the four different compactness grades.\label{VisplotGrades}}                                   
\end{figure}

Figure \ref{VisplotGrades} shows examples of the visibility plots
for sources of each grade, A through D.                                            
Since we could not reliably detect sources with correlated flux density
$<30$ mJy, it was not always possible to measure weak sources $<100$ mJy
with a normalized fringe visibility of 0.2.   
 
Due to overlapping observing fields around the target masers, some
sources were observed multiple times. Since the observations
took place at different times, the
fitted coordinates, uncertainties, and assigned grades can differ
among observations. 
For sources with grades A through D that were observed two or more times, we determined
weighted means and standard deviations for the coordinates
in Right Ascension and Declination. The formal position variances from the
delay and rate fits served as the weights.  To guard against unrealistically 
small variances, which can come from limited data, we established
a lower limit of $0.003\arcsec$ for individual position uncertainties. 
We found the rms position deviations were generally larger than 
that expected from the formal position uncertainties,
indicating the levels of systematic errors present in
our ``snap-shot'' VLBA observations.
 
While most sources that were observed multiple times had the same compactness grade,
some borderline cases were found that had mixed grades.
Therefore, we assigned numerical grades of 4 through 1 for
grades A through D, and obtained an (unweighted) average source grade.
The multiple observed sources were then placed in four groups according to their average
grade.  The grade A group contained all sources with numerical grades
above 3.5, the grade B group between 2.5 and 3.5, the grade C group 
between 1.5 and 2.5, and the grade D group below 1.5. 17\% of the observed ICRF2 sources
have a bad grade (C or D).
 
Finally, we calculated the average of the individual position 
standard deviations for all sources in each group. 
In order to avoid the rare very poor measurement from biasing 
the average deviation, we discarded individual sources with
position standard deviations exceeding 50 mas.
The average standard deviations (global grade errors) serve as
empirical indications of true source position uncertainties and 
are summarized in Table \ref{Statistics}.
 
The empirically determined position errors for each grade were then used 
as an ``error floor'' for the position errors reported in Table \ref{DetecSour}. 
Since, in this first-pass processing potentially unreliable formal errors were used 
for weighting, we recalculated the mean coordinates and the standard errors of the
mean with data weights formed from the geometric mean of the formal fitting errors 
and the error floors for the appropriate source grade. 
For the 30 sources in common with the \citet{Petrov2011} survey, their positions
agree within the joint uncertainties. A comparison with the ICRF2 catalog is not meaningful 
for the 25 sources whose positions have been improved by averaging the measured and the ICRF2 coordinates.
The coordinates of the remaining 45 ICRF2 sources coincide with the ICRF2 catalog positions 
within the joint uncertainties.
 
\begin{figure}
	\centering	
		\includegraphics[width=8cm]{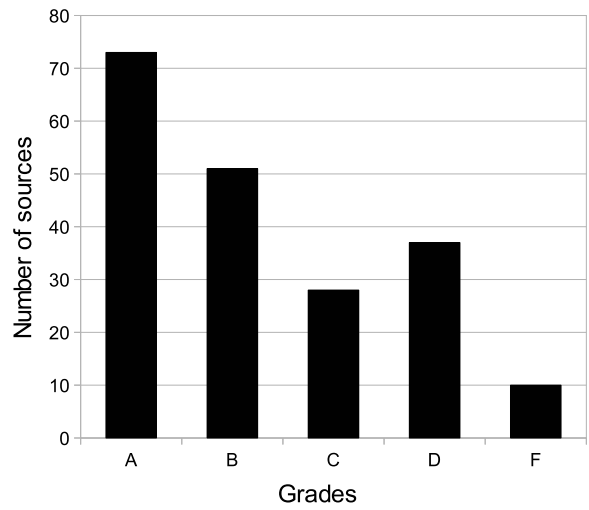}
			\caption{Histogram of the distribution of sources by compactness grade.\label{Histogram_Grades}}
\end{figure}

Table \ref{DetecSour} lists all detected sources together
with the estimated coordinates (Column 2,3), position
uncertainties (Column 4,5), grades (Column 6), fluxes (Column 7), 
and the number of observations (Column 8).  This table also contains the grade
F sources for which the coordinates come
from the NVSS catalog. The histogram in Fig. \ref{Histogram_Grades} 
shows the distribution of the sources over the four grade groups.

\section{Non-Detections} 
 
A total of 1330 objects were not detected with the VLBA. Their coordinates, position uncertainties,
and flux densities are extracted from the NVSS and CORNISH point source catalogs (see Table \ref{UndetecSour}, 
the entire table will be published in the online version).
The non-detected sources could be galactic (e.g. HII regions, SNRs) or extragalactic (e.g. lobes of radio sources). 
A comparison with catalogs of Galactic HII regions 
\citep{Kuchar1997,Paladini2003,Quireza2006}
yields less than 50 possible identifications. 
In order to determine if the
remaining objects are likely extragalactic, we estimated the number
of detectable extragalactic sources from extragalactic source counts.
Since most of the non-detected sources were selected from the NVSS
catalog at 1.4 GHz, we used a normalized differential source count
at this frequency provided by \citet{Condon1984}.                     
Since most of the sources in our non-detection list have flux densities
above $\sim50$ mJy, we chose this value as a lower limit for the
source count calculations. Integrating the source number function
as a function of flux density from 50 mJy to $\sim1$ Jy yields
a predicted value of $\sim12000$ radio sources per steradian. Since our survey
covered an area of $\approx600$ square degrees, one would have expected 
$\sim1300$ detections, which is consistent with our observations.
Thus, we assume that most of the non-detected sources are 
extragalactic.  

\section{Concluding Remarks}

In this paper, we present the results of a calibrator search for the BeSSeL survey. 
Of the 1529 radio continuum sources observed, 199 candidate calibrator sources were detected.
Most of the 1330 undetected sources are probably of extragalactic origin.
The list of calibrator candidates will be made available on the homepage
of the BeSSeL survey. Furthermore, the ALMA and EVLA communities can use this list as a starting point
for their own calibrator surveys.

\clearpage

\begin{table} 
\begin{center}
\caption{Results of the position error analysis of the detected sources.\label{Statistics}} 
\begin{tabular}{ccrcr}
\tableline\tableline 
			Grade & No. calibrators with & $\sigma_{R.A. cos(Dec.)}$ & No. Calibrators
with & $\sigma_{Dec.}$\\                                                
                & $\sigma_{R.A. cos(Dec.)}$ < 50 mas  &   [mas] &
$\sigma_{Dec.}$ < 50 mas  &  [mas]\\ \tableline                
    A & 23 &  1.60 & 23 &  2.68\\
    B & 13 &  2.22 & 13 &  5.07\\
    C & 13 &  8.64 & 13 & 11.89\\
    D & 11 & 11.42 & 11 & 13.54\\
\tableline
\end{tabular}
\tablecomments{Column 1 gives the compactness grade described in the text.
Columns 2 and 4 give the number of calibrators (per grade) with position standard 
deviations smaller than 50 mas. Column 3 and 5 contain the empirically estimated 
uncertainties in Right Ascension and Declination for each grade.} 
\end{center}
\end{table}

\begin{deluxetable}{lccrrccc}
\centering
\tabletypesize{\small}
\tablecaption{\label{DetecSour} Detected sources}
\tablehead{                       
			\colhead{Name} & \colhead{R.A.}  & \colhead{Dec.} & \colhead{1$\sigma$-Unc}
& \colhead{1$\sigma$-Unc} & \colhead{Grade} &  \colhead{Flux} &  \colhead{No.} \\                          
	\colhead{}		&\colhead{(J2000)} &\colhead{(J2000)} & \colhead{R.A. $\cos$(Dec.)} & \colhead{Dec.} &\colhead{} &\colhead{ }& \colhead{Obs}\\ & \colhead{(hh mn ss.sssss)}  & \colhead{(dd $\arcmin\arcmin$ $\arcsec\arcsec.\arcsec\arcsec\arcsec\arcsec$)}
&   \colhead{(mas)} &  \colhead{(mas)} &\colhead{} & \colhead{(Jy)} & }                                     
\tablecomments{Columns 1 through 3 give the
source name and its Right Ascension (R.A.) and Declination (Dec.). The asteriks mark
sources for which the ICRF2 source position was averaged in with the source position.  
Weighted means are given for those calibrators that were observed
more than once. Columns 4 and 5 give the coordinate uncertainties, determined
from the quadrature sum of the formal fitting errors and the error floors for
the appropriate compactness grade.
For the grade F sources, the coordinates from the NVSS catalog were used.
Column 6 and 7 contain the compactness grades, A through F, as well as the source fluxes. 
Column 8 gives the number of observations for each source.}
\startdata
J0012+6551 & 00 12 37.67279 & +65 51 10.8236 & 12.1 & 14.1 & D &  0.20 &  1\\
J0035+6130 & 00 35 25.31071 & +61 30 30.7637 & 3.4 &    4.0 & A &  0.23 &  1\\
J0037+6236 & 00 37 04.33092 & +62 36 33.3045 & 4.6 &    6.5 & B &  0.25 &  1\\
J0244+6228 & 02 44 57.69662 & +62 28 06.5172 & 3.7 &    5.9 & B &  2.00 &  1\\
J0248+6214 & 02 48 58.89198 & +62 14 09.6783 & 4.6 &    6.5 & B &  0.10 &  1\\
J0251+6824 & 02 51 45.79153 & +68 24 01.7686 & 5.2 &    5.7 & A &  0.06 &  1\\
J0256+6827 & 02 56 10.82729 & +68 27 51.0042 & 5.2 &    5.7 & A &  0.07 &  1\\
J0304+6821 & 03 04 22.00321 & +68 21 37.4748 & 10.1 &   10.4 & A &  0.60 &  1\\
J0306+6243 & 03 06 42.65955 & +62 43 02.0244 &    3.4 &    4.0 & A &  0.40 &  1\\
J0308+6955 & 03 08 27.81517 & +69 55 58.9545 &   10.2 &   11.2 & B &  0.07 &  1\\
J0455+4653 & 04 55 53.20678 & +46 53 30.2520 &    9.3 &   10.3 & B &  0.05 &  1\\
J0455+4803 & 04 55 07.17204 & +48 03 47.3791 &    3.7 &    5.9 & B &  0.07 &  1\\
J0502+4729 & 05 02 22.33186 & +47 29 12.0753 &    8.2 &    8.4 & A &  0.06 &  1\\
J0507+4645 & 05 07 23.65891 & +46 45 42.3372 &    4.3 &    4.8 & A &  0.30 &  1\\
J0507+4720 & 05 07 20.43482 & +47 20 36.3146 &    5.2 &    5.7 & A &  0.08 &  1\\
J0509+3951 & 05 09 48.81719 & +39 51 54.6176 &    7.2 &    7.5 & A &  0.25 &  1\\
J0512+4041 & 05 12 52.54295 & +40 41 43.6188 &    3.4 &    4.0 & A &  1.00 &  1\\
J0514+4025 & 05 14 02.81508 & +40 25 24.0477 &   10.5 &   13.3 & C &  0.08 &  1\\
J0523+3926 & 05 23 51.23640 & +39 26 57.7358 &    3.7 &    5.9 & B &  0.06 &  1\\
J0529+3209 & 05 29 28.21154 & +32 09 46.9452 &    5.2 &    5.7 & A &  0.08 &  1\\
J0535+3249 & 05 35 52.08521 & +32 49 17.3790 &   14.1 &   14.3 & A &  0.05 &  1\\
J0539+3308 & 05 39 09.67217 & +33 08 15.4957 &    3.7 &    5.9 & B &  0.06 &  1\\
J0540+2507 & 05 40 14.34318 & +25 07 55.3565 &    3.3 &    3.7 & A &  0.09 &  2\\
J0541+3126 & 05 41 35.23810 & +31 26 44.2135 &   21.1 &   21.6 & B &  0.05 &  1\\
J0541+3301 & 05 41 49.43590 & +33 01 31.8900 &    7.2 &    7.5 & A &  0.08 &  1\\
J0545+3217 & 05 45 16.48160 & +32 17 27.0506 &    5.2 &    5.7 & A &  0.07 &  1\\
J0549+3054 & 05 49 54.18174 & +30 54 47.5885 &    6.2 &    6.6 & A &  0.08 &  1\\
J0550+2326 & 05 50 47.39145 & +23 26 48.1628 &    6.2 &    6.6 & A &  0.26 &  1\\
J0557+2413 & 05 57 04.71415 & +24 13 55.2844 &    6.2 &    6.6 & A &  1.00 &  1\\
J0603+2159 & 06 03 51.55708 & +21 59 37.6982 &    3.7 &    5.9 & B &  0.08 &  1\\
J0607+2129 & 06 07 59.56570 & +21 29 43.7200 &    3.7 &    5.9 & B &  0.01 &  1\\
J0607+2218 & 06 07 17.43600 & +22 18 19.0800 &    3.7 &    5.9 & B &  0.01 &  1\\
J0608+2229 & 06 08 34.31090 & +22 29 42.9810 &    3.7 &    5.9 & B &  0.06 &  1\\
J0649+0008 & 06 49 16.33743 & +00 08 56.0797 &    3.4 &    4.0 & A &  0.13 &  1\\
J0650+0358 & 06 50 38.13429 & +03 58 08.4420 &    3.4 &    4.0 & A &  0.15 &  1\\
J0656+0315 & 06 56 59.18022 & +03 15 53.4395 &    4.6 &    6.5 & B &  0.05 &  1\\
J0657+0224 & 06 57 12.36716 & +02 24 32.7362 &    3.4 &    4.0 & A &  0.09 &  1\\
J0703--0051 & 07 03 19.08675 & --00 51 03.1574 &    3.4 &    4.0 & A &  0.50 &  1\\
J0721--1530 & 07 21 13.49140 & --15 30 41.0088 &    4.3 &    4.8 & A &  0.15 &  1\\
J0721--1630 & 07 21 49.13771 & --16 30 19.7464 &    6.2 &    6.6 & A &  0.12 &  1\\
J0724--1545 & 07 24 59.00632 & --15 45 29.3698 &    8.2 &    8.4 & A &  0.08 &  1\\
J0729--1320 & 07 29 17.81773 & --13 20 02.2600 &    3.4 &    4.0 & A &  0.15 &  1\\
J0735--1735 & 07 35 45.81270 & --17 35 48.5023 &    3.4 &    5.0 & B &  0.29 &  2\\
J0738--2403 & 07 38 09.99020 & --24 03 56.3766 &    6.2 &    6.6 & A &  0.07 &  1\\
J0739--2301 & 07 39 24.99814 & --23 01 31.8848 &    4.3 &    4.8 & A &  0.08 &  1\\
J0740--2444 & 07 40 14.71735 & --24 44 36.6991 &    4.6 &    6.5 & B &  0.40 &  1\\
J0741--1937 & 07 41 52.78735 & --19 37 34.8280 &    3.4 &    4.0 & A &  0.06 &  1\\
J0742--2043 & 07 42 32.10076 & --20 43 41.2244 &   21.8 &   23.3 & C &  0.07 &  1\\
J0745--1828 & 07 45 19.43909 & --18 28 24.7985 &    3.4 &    4.0 & A &  0.08 &  1\\
J0745--2451 & 07 45 10.26453 & --24 51 43.7704 &    4.3 &    4.8 & A &  0.60 &  1\\
J0749--2344 & 07 49 51.77926 & --23 44 48.7882 &    3.4 &    4.0 & A &  0.12 &  1\\
J0750--2451 & 07 50 00.40201 & --24 51 36.6085 &    3.4 &    4.0 & A &  0.09 &  1\\
J1712--3514 & 17 12 05.13526 & --35 14 34.1834 &   23.0 &   24.2 & D &  0.09 &  1\\
J1712--3736 & 17 12 38.79641 & --37 36 46.7341 &   13.2 &   15.5 & C &  0.18 &  1\\
J1717--3342* & 17 17 36.03045 & --33 42 08.7942 &    7.8 &    9.6 & D &  1.40 &  2\\
J1717--3948 & 17 17 38.59784 & --39 48 52.5572 &   18.9 &   20.2 & D &  0.15 &  1\\
J1719--3354 & 17 19 34.35417 & --33 54 48.8743 & 1000.1 & 1000.1 & D &  0.08 &  1\\
J1723--3936 & 17 23 32.18591 & --39 36 41.9943 &   11.8 &   13.9 & D &  0.09 &  1\\
J1733--3722* & 17 33 15.19300 & --37 22 32.3955 &    2.4 &    2.8 & A &  2.00 &  2\\
J1743--3058 & 17 43 17.88572 & --30 58 18.6690 &   23.0 &   24.2 & D &  0.14 &  1\\
J1751--1950 & 17 51 41.34380 & --19 50 47.5060 &    2.5 &    3.3 & B &  0.08 &  2\\
J1751--2523 & 17 51 51.26230 & --25 24 00.0730 &   13.2 &   15.5 & C &  0.23 &  1\\
J1751--2524* & 17 51 51.26204 & --25 24 00.0588 &   10.8 &   12.6 & C &  0.25 &  2\\
J1752--2956 & 17 52 33.10790 & --29 56 44.8983 &   18.2 &   19.9 & C &  0.07 &  1\\
J1752--3001 & 17 52 30.95041 & --30 01 06.6526 &   15.2 &   15.8 & B &  0.14 &  1\\
J1755--2232* & 17 55 26.28462 & --22 32 10.6128 &    5.6 &    6.5 & C &  0.25 &  4\\
J1758--2343 & 17 58 23.01669 & --23 43 12.1257 &    7.7 &    8.9 & D &  0.25 &  3\\
J1800--2107 & 18 00 44.51594 & --21 07 33.7637 &  170.4 &  170.5 & D &  0.05 &  1\\
J1801--2214 & 18 01 43.54953 & --22 14 28.8094 &   10.8 &   11.5 & D &  0.10 &  4\\
J1805--2512 & 18 05 23.54818 & --25 12 38.7469 &    9.1 &   12.3 & C &  0.08 &  1\\
J1807--2506* & 18 07 40.68757 & --25 06 25.9440 &    4.0 &    5.2 & A &  0.23 &  3\\
J1808--1822 & 18 08 55.51552 & --18 22 53.3897 &    3.9 &    4.8 & C &  0.08 &  7\\
J1809--1520 & 18 09 10.20936 & --15 20 09.6991 &    3.4 &    4.0 & A &  0.07 &  1\\
J1809--1546 & 18 09 09.25000 & --15 46 53.8000 &  \nodata &  \nodata & F &  \nodata &  1\\
J1810--1626 & 18 10 39.85118 & --16 26 52.9225 &    5.6 &    6.9 & C &  0.08 &  5\\
J1811--2055 & 18 11 06.79427 & --20 55 03.2781 &    7.8 &    8.9 & D &  0.15 &  4\\
J1813--2015 & 18 13 16.87269 & --20 15 44.0791 &   10.7 &   11.9 & D &  0.05 &  2\\
J1815--1836 & 18 15 30.46000 & --18 36 14.0000 &  \nodata &  \nodata & F &  \nodata &  1\\
J1818--1705 & 18 18 02.90348 & --17 05 40.8859 &    9.0 &   10.1 & D &  0.16 &  3\\
J1819--2036 & 18 19 36.89582 & --20 36 31.5519 &    8.6 &   10.4 & D &  0.09 &  2\\
J1821--0502* & 18 21 11.80930 & --05 02 20.0900 &    6.4 &    7.9 & B &  0.20 &  1\\
J1821--1224 & 18 21 23.27826 & --12 24 12.9142 &   10.5 &   13.3 & C &  0.04 &  1\\
J1821--2110 & 18 21 05.46937 & --21 10 45.2453 &    4.2 &    5.8 & C &  0.12 &  2\\
J1825--0737* & 18 25 37.60958 & --07 37 30.0129 &    2.5 &    3.3 & B &  0.08 &  2\\
J1825--1551 & 18 25 11.72184 & --15 51 34.4066 &   13.2 &   15.5 & C &  0.08 &  1\\
J1825--1718* & 18 25 36.53225 & --17 18 49.8474 &    1.3 &    2.1 & B &  0.30 &  8\\
J1827--0405* & 18 27 45.04040 & --04 05 44.5800 &    3.7 &    5.9 & B &  0.13 &  1\\
J1828--0530* & 18 28 40.15371 & --05 30 50.8918 &    9.1 &   12.3 & C &  0.07 &  1\\
J1828--0912 & 18 28 56.02434 & --09 12 31.1181 &   15.9 &   17.4 & D &  0.07 &  1\\
J1832--0610 & 18 32 42.23000 & --06 10 26.5000 &  \nodata &  \nodata & F &  \nodata &  1\\
J1832--1035 & 18 32 20.83656 & --10 35 11.2006 &   15.2 &   16.8 & D & \nodata &  1\\
J1832--2039* & 18 32 11.04646 & --20 39 48.1974 &    2.5 &    3.3 & B &  0.50 &  2\\
J1833--0323 & 18 33 23.90553 & --03 23 31.4656 &    6.4 &    7.9 & B &  0.12 &  1\\
J1833--0855 & 18 33 19.58200 & --08 55 27.2100 &   15.2 &   16.8 & D & \nodata &  1\\
J1834--0301* & 18 34 14.07459 & --03 01 19.6268 &    1.6 &    2.5 & B &  0.20 &  7\\
J1837--0653 & 18 37 58.03336 & --06 53 31.2470 &    9.8 &   11.1 & D &  0.07 &  2\\
J1841--0348 & 18 41 27.31540 & --03 48 44.3323 &    7.5 &    8.5 & D &  0.10 &  5\\
J1846--0003 & 18 46 03.78257 & --00 03 38.2834 &    8.0 &    9.7 & C &  0.10 &  3\\
J1846--0651* & 18 46 06.30024 & --06 51 27.7472 &    2.9 &    4.3 & B &  0.11 &  2\\
J1847--0004 & 18 47 04.96160 & --00 04 47.2000 &   20.1 &   21.0 & D &  0.05 &  2\\
J1848+0138 & 18 48 21.81035 & +01 38 26.6322 &    3.7 &    5.9 & B &  0.12 &  1\\
J1849+0349 & 18 49 29.17000 & +03 49 55.0000 &  \nodata &  \nodata & F &  \nodata &  1\\
J1851+0035 & 18 51 46.72295 & +00 35 32.3649 &    5.9 &    6.9 & D &  0.07 &  4\\
J1852+0308 & 18 52 37.01000 & +03 08 38.9000 &  \nodata &  \nodata & F &  \nodata &  1\\
J1853--0010 & 18 53 10.26917 & --00 10 50.7398 &    9.3 &   10.7 & D &  0.07 &  3\\
J1853--0048 & 18 53 41.98921 & --00 48 54.3307 &    5.4 &    7.2 & C &  0.08 &  3\\
J1854+0542 & 18 54 36.12843 & +05 42 59.3068 &    9.5 &   12.5 & C &  0.10 &  1\\
J1855+0215 & 18 55 00.11300 & +02 15 41.1000 &  100.7 &  100.9 & D &  0.03 &  1\\
J1855+0251* & 18 55 35.43646 & +02 51 19.5629 &    2.4 &    3.8 & C &  0.12 &  4\\
J1856+0610* & 18 56 31.83900 & +06 10 16.7678 &    3.5 &    5.3 & C &  0.25 &  2\\
J1857+0051 & 18 57 49.55000 & +00 51 19.2000 &  \nodata &  \nodata & F &  \nodata &  1\\
J1857--0048 & 18 57 51.35860 & --00 48 21.9496 &    3.4 &    4.0 & A &  0.11 &  1\\
J1858+0313 & 18 58 02.35272 & +03 13 16.3116 &   12.1 &   13.3 & D &  0.70 &  2\\
J1900+0003 & 19 00 17.57016 & +00 03 55.9078 &   13.4 &   15.2 & D &  0.07 &  1\\
J1900+0303 & 19 00 33.59944 & +03 03 46.0646 &   11.3 &   12.9 & D &  0.14 &  2\\
J1900--0009 & 19 00 49.56848 & --00 09 15.3823 &   50.7 &   51.4 & C &  0.05 &  1\\
J1903+0145 & 19 03 53.06326 & +01 45 26.3108 &    3.2 &    3.8 & B &  0.18 &  2\\
J1904+0110 & 19 04 26.39738 & +01 10 36.7078 &    3.2 &    3.8 & B &  0.12 &  2\\
J1905+0952 & 19 05 39.89889 & +09 52 08.4071 &    1.8 &    2.1 & A &  0.25 &  4\\
J1906+0139 & 19 06 12.96214 & +01 39 13.6688 &   15.2 &   16.8 & D &  0.10 &  1\\
J1907+0127 & 19 07 11.99613 & +01 27 08.9644 &    3.0 &    3.6 & B &  0.25 &  3\\
J1907+0907 & 19 07 41.96336 & +09 07 12.3968 &    3.3 &    4.6 & B &  0.25 &  2\\
J1908+1201 & 19 08 40.32064 & +12 01 58.8609 &    3.4 &    4.0 & A &  0.11 &  1\\
J1911+1611 & 19 11 58.25738 & +16 11 46.8600 &    3.4 &    4.0 & A &  0.50 &  1\\
J1913+0932 & 19 13 24.02539 & +09 32 45.3791 &   10.0 &   12.9 & C &  0.16 &  1\\
J1913+1307 & 19 13 14.00638 & +13 07 47.3307 &    3.4 &    4.0 & A &  0.08 &  1\\
J1917+1405 & 19 17 18.06412 & +14 05 09.7675 &    3.3 &    3.8 & B &  0.12 &  2\\
J1921+1625 & 19 21 57.38244 & +16 25 01.9231 &    3.4 &    4.0 & A &  0.19 &  1\\
J1922+0841* & 19 22 18.63360 & +08 41 57.3780 &    5.2 &    5.7 & A &  0.10 &  1\\
J1922+1504 & 19 22 33.27275 & +15 04 47.5388 &    9.1 &   12.3 & D &  0.08 &  2\\
J1922+1530 & 19 22 34.69942 & +15 30 10.0340 &    6.4 &    8.5 & C &  0.40 &  3\\
J1924+1540 & 19 24 39.45589 & +15 40 43.9380 &    2.3 &    2.8 & B &  0.50 &  3\\
J1925+1227* & 19 25 40.81708 & +12 27 38.0856 &    1.7 &    2.0 & A &  0.14 &  5\\
J1927+1847 & 19 27 32.31207 & +18 47 07.9048 &    3.4 &    4.0 & A &  0.07 &  1\\
J1928+1859 & 19 28 42.71875 & +18 59 24.5627 &    5.5 &    7.1 & B &  0.07 &  1\\
J1930+1532 & 19 30 52.76690 & +15 32 34.4285 &    3.4 &    4.0 & A &  0.45 &  1\\
J1934+1043* & 19 34 35.02560 & +10 43 40.3660 &    3.4 &    4.0 & A &  0.08 &  1\\
J1936+2246 & 19 36 29.30467 & +22 46 25.8540 &    3.4 &    4.0 & A &  0.08 &  1\\
J1936+2357 & 19 36 00.92503 & +23 57 31.9722 &    3.4 &    4.0 & A &  0.20 &  1\\
J1938+2348 & 19 38 45.60688 & +23 48 15.7092 &   15.2 &   16.8 & D &  0.06 &  1\\
J1941+2307 & 19 41 55.11140 & +23 07 56.5250 &    3.7 &    5.9 & B &  0.01 &  1\\
J1946+2255 & 19 46 22.29846 & +22 55 24.6282 &    8.3 &    9.5 & B &  0.08 &  1\\
J1946+2300 & 19 46 06.25140 & +23 00 04.4145 &    3.7 &    5.9 & B &  0.07 &  1\\
J1946+2418 & 19 46 19.96067 & +24 18 56.9093 &   50.0 &   50.3 & B &  0.06 &  1\\
J1955+3233 & 19 55 56.35092 & +32 33 04.5060 &   11.8 &   14.3 & C &  0.07 &  1\\
J1957+3216 & 19 57 43.00353 & +32 16 33.5832 &    3.4 &    4.0 & A &  0.07 &  1\\
J1957+3338 & 19 57 40.54974 & +33 38 27.9429 &    5.2 &    5.7 & A &  0.16 &  1\\
J1957+3427 & 19 57 34.44939 & +34 27 54.6774 &  100.7 &  100.9 & D &  0.05 &  1\\
J2001+3216 & 20 01 16.77409 & +32 16 46.9435 &    4.3 &    4.8 & A &  0.06 &  1\\
J2001+3323 & 20 01 42.20694 & +33 23 44.7461 &   10.5 &   13.3 & C &  0.30 &  1\\
J2007+4029 & 20 07 44.94494 & +40 29 48.6009 &    3.5 &    5.2 & B &  3.50 &  2\\
J2009+3543 & 20 09 57.63785 & +35 43 18.0079 &   11.8 &   14.3 & C &  0.33 &  1\\
J2010+3322 & 20 10 49.72256 & +33 22 13.8074 &   13.2 &   15.5 & C &  0.80 &  1\\
J2015+3410 & 20 15 28.83206 & +34 10 39.4175 &    3.7 &    5.9 & B &  0.40 &  1\\
J2015+3710 & 20 15 28.72950 & +37 10 59.5179 &    2.4 &    2.8 & A &  2.40 &  2\\
J2016+3600 & 20 16 45.61840 & +36 00 33.3774 &    3.7 &    5.9 & B &  0.16 &  1\\
J2018+3812 & 20 18 42.84963 & +38 12 41.3396 &   12.1 &   14.1 & D &  0.08 &  1\\
J2018+3851 & 20 18 31.02410 & +38 51 19.3854 &    3.7 &    5.9 & B &  0.33 &  1\\
J2020+4057 & 20 20 36.17955 & +40 57 53.0792 &   19.7 &   21.0 & D &  0.10 &  1\\
J2025+3343 & 20 25 10.84328 & +33 43 00.1955 &   12.2 &   13.0 & B &  4.50 &  1\\
J2025+4335 & 20 25 18.96000 & +43 35 24.0000 &  \nodata &  \nodata & F &  \nodata &  1\\
J2027+3612 & 20 27 22.15431 & +36 12 14.0702 &   10.5 &   12.0 & D &  0.12 &  2\\
J2030+3700 & 20 30 31.26724 & +37 00 36.0379 &   12.5 &   14.4 & D &  0.20 &  1\\
J2032+4057 & 20 32 25.77049 & +40 57 27.9154 &   10.7 &   12.2 & D &  0.10 &  2\\
J2033+4000 & 20 33 03.66147 & +40 00 24.3573 &   11.8 &   13.9 & D &  0.09 &  1\\
J2038+5119 & 20 38 37.03480 & +51 19 12.6592 &    3.1 &    3.6 & A &  1.80 &  2\\
J2056+4940 & 20 56 42.73987 & +49 40 06.6010 &    3.4 &    3.9 & A &  0.09 &  2\\
J2059+4851 & 20 59 57.87249 & +48 51 12.7002 &    4.3 &    4.8 & A &  0.10 &  1\\
J2100+5331 & 21 00 56.33534 & +53 31 33.1413 &    3.7 &    5.9 & B &  0.09 &  1\\
J2102+4702* & 21 02 17.05620 & +47 02 16.2530 &    3.4 &    4.0 & A &  0.17 &  1\\
J2105+5356 & 21 05 12.79000 & +53 56 09.4000 &  \nodata &  \nodata & F &  \nodata &  1\\
J2114+4953 & 21 14 15.03646 & +49 53 40.9540 &    4.0 &    4.4 & A &  0.12 &  2\\
J2117+5431* & 21 17 56.48434 & +54 31 32.5036 &    2.4 &    2.8 & A &  0.25 &  2\\
J2123+5452 & 21 23 46.83487 & +54 52 43.4884 &    4.3 &    4.8 & A &  0.09 &  1\\
J2123+5500* & 21 23 05.31345 & +55 00 27.3253 &    2.4 &    2.8 & A &  0.25 &  2\\
J2125+6423* & 21 25 27.44700 & +64 23 39.3550 &    3.4 &    4.0 & A &  0.35 &  1\\
J2127+5528 & 21 27 32.27527 & +55 28 33.9771 &    9.1 &   12.3 & C &  0.09 &  1\\
J2131+5214 & 21 31 58.22000 & +52 14 15.9000 &  \nodata &  \nodata & F &  \nodata &  1\\
J2137+5101 & 21 37 00.98625 & +51 01 36.1292 &    2.4 &    2.8 & A &  0.35 &  2\\
J2139+5300 & 21 39 53.62438 & +53 00 16.5993 &    3.7 &    5.9 & B &  0.12 &  1\\
J2139+5540 & 21 39 32.61754 & +55 40 31.7711 &   12.9 &   14.8 & D &  0.05 &  1\\
J2145+5147 & 21 45 07.66657 & +51 47 02.2430 &    2.5 &    3.3 & B &  0.06 &  2\\
J2148+6107 & 21 48 16.04239 & +61 07 05.7966 &    3.7 &    5.9 & B &  0.60 &  1\\
J2150+5103 & 21 50 14.26619 & +51 03 32.2638 &    2.4 &    2.8 & A &  0.14 &  2\\
J2159+6606 & 21 59 24.68626 & +66 06 52.2906 &    3.4 &    4.0 & A &  0.18 &  1\\
J2203+6750* & 22 03 12.62260 & +67 50 47.6730 &    3.4 &    4.0 & A &  0.22 &  1\\
J2209+5158* & 22 09 21.48690 & +51 58 01.8330 &    3.4 &    4.0 & A &  0.18 &  1\\
J2214+5250 & 22 14 55.76000 & +52 50 58.2000 &  \nodata &  \nodata & F &  \nodata &  1\\
J2217+5202 & 22 17 54.46074 & +52 02 51.3698 &    4.3 &    4.8 & A &  0.06 &  1\\
J2225+5120 & 22 25 25.39414 & +51 20 45.7403 &   20.1 &   20.6 & B &  0.08 &  1\\
J2243+6055 & 22 43 00.81457 & +60 55 44.2156 &    3.1 &    3.7 & B &  0.06 &  2\\
J2250+5550 & 22 50 42.85149 & +55 50 14.5730 &    9.1 &    9.4 & A &  0.35 &  1\\
J2254+6209 & 22 54 25.29266 & +62 09 38.7227 &    2.4 &    3.6 & B &  0.08 &  4\\
J2257+5720 & 22 57 22.04611 & +57 20 30.1966 &   11.1 &   11.3 & A &  0.15 &  1\\
J2258+5719 & 22 58 57.94146 & +57 19 06.4599 &    6.8 &    7.0 & A &  0.18 &  2\\
J2301+5706 & 23 01 26.62714 & +57 06 25.4987 &   10.1 &   10.4 & A &  0.12 &  1\\
J2302+6405 & 23 02 41.31514 & +64 05 52.8506 &    3.0 &    4.1 & B &  0.12 &  2\\
J2314+5813 & 23 14 19.08325 & +58 13 47.6474 &    9.1 &   12.3 & C &  0.07 &  1\\
J2339+6010* & 23 39 21.12514 & +60 10 11.8504 &    2.4 &    2.8 & A &  0.30 &  2\\
\enddata
\end{deluxetable} 
 
\begin{table} 
\begin{center}
\caption{Undetected sources.\label{UndetecSour}} 
\begin{tabular}{ccccc}
\tableline\tableline 
Name & R.A. (J2000) & Dec. (J2000) & Unc & Flux\\
     & (hh mm ss.ss)&(dd $\arcmin\arcmin$ $\arcsec\arcsec.\arcsec$) & ($\arcsec$) & Jy\\\tableline
J0022+6351 & 00 22 52.31 & +63 51 21.8 & 0.7 & 0.030 \\
J0023+6349 & 00 23 25.39 & +63 49 45.5 & 0.9 & 0.022 \\
J0027+6425 & 00 27 34.96 & +64 25 55.1 & 0.7 & 0.041 \\
J0028+6227 & 00 28 27.82 & +62 27 12.8 & 0.6 & 0.148 \\
J0028+6345 & 00 28 40.03 & +63 45 59.9 & 0.6 & 0.074 \\
J0029+6303 & 00 29 23.38 & +63 03 35.5 & 0.6 & 0.096 \\
J0029+6358 & 00 29 45.11 & +63 58 41.5 & 0.6 & 2.992 \\
J0031+6322 & 00 31 32.49 & +63 22 08.5 & 0.6 & 0.108 \\
J0032+6301 & 00 32 11.97 & +63 01 00.7 & 0.7 & 0.041 \\
J0033+6428 & 00 33 38.87 & +64 28 45.4 & 0.6 & 0.107 \\
\tableline
\end{tabular}
\tablecomments{The coordinates, coordinate
uncertainties, and the fluxes are extracted from the CORNISH and NVSS
point source catalogs. The entire table is available in the online version.}
\end{center}
\end{table} 
 
\end{document}